\documentstyle[epsfig,12pt]{article}

\normalsize
\def\st{\ifmmode{\tilde{t}} \else{$\tilde{t}$} \fi}
\def\sb{\ifmmode{\tilde{b}} \else{$\tilde{b}$} \fi}
\def\sq{\ifmmode{\tilde{q}} \else{$\tilde{q}$} \fi}
\def\sg{\ifmmode{\tilde{g}} \else{$\tilde{g}$} \fi}
\def\sz{\ifmmode{\tilde{\chi}^0} \else{$\tilde{\chi}^0$} \fi}
\def\sw{\ifmmode{\tilde{\chi}} \else{$\tilde{\chi}$} \fi}
\def\sl{\ifmmode{\tilde{\ell}} \else{$\tilde{\ell}$} \fi}
\def\sn{\ifmmode{\tilde{\nu}} \else{$\tilde{\nu}$} \fi}
\def\stau{\ifmmode{\tilde{\tau}} \else{$\tilde{\tau}$} \fi}
\def\nle{\rlap{\lower 3.5 pt \hbox{$\mathchar \sim$}}\raise 1pt \hbox{$<$}}
\def\nge{\rlap{\lower 3.5 pt \hbox{$\mathchar \sim$}}\raise 1pt \hbox{$>$}}

\begin{document}

\vspace*{-0.8cm}
\begin{flushright}
hep-ph/9709253
\end{flushright}

\vspace{0.3cm}

\begin{center}
{\LARGE \bf 
Higgs Particle Decays in Supersymmetry\footnote{
To appear in  {\it Proceedings of ``Beyond the Standard
Model V'', Balholm, Norway, April 29 - May 4, 1997}}
}
\end{center}

\vspace{3mm}

\noindent
\large{A.~Bartl$^a$, \underline{H.~Eberl}$^b$,
K.~Hidaka$^c$, T.~Kon$^d$, W.~Majerotto$^b$, and Y.~Yamada$^e$}\\

\noindent\small{\it
$^a$ Institut f\"ur Theoretische Physik, Universit\"at Wien,
A-1090 Vienna, Austria\\
$^b$ Institut f\"ur Hochenergiephysik der \"Osterreichischen
Akademie der Wissenschaften,\\ A-1050 Vienna, Austria\\
$^c$ Department of Physics, Tokyo Gakugei University, Koganei,
Tokyo 184, Japan\\
$^d$ Faculty of Engineering, Seikei University, Musashino,
Tokyo 180, Japan\\
$^e$ Department of Physics, Tohoku University, Sendai 980-77, Japan
}

\vspace{5mm}

\begin{abstract}
We present a detailed study of the decays of the Higgs bosons $H^+$,
$H^0$, and $A^0$ within the Minimal Supersymmetric Standard Model
including SUSY--QCD corrections. We find that the supersymmetric modes
$\tilde t \bar{\tilde b}$ ($\tilde t \bar{\tilde t}$, and for large $\tan\beta$
$\tilde b \bar{\tilde b}$) can dominate the $H^+$ ($H^0$, $A^0$) decays
in a wide range of the model parameters due to the large Yukawa couplings
and mixings of $\tilde t$ and $\tilde b$. Compared to the
conventional modes $H^+ \to \tau^+ \nu_{\tau}, t \bar b$, and $H^0, A^0 \to
t \bar t, b \bar b$, the supersymmetric modes can have an important impact on
the Higgs boson searches at future colliders.
\end{abstract}

\section{Introduction}
The Minimal Supersymmetric Standard Model (MSSM)
\cite{eberl:cit1} implies the existence of
five physical Higgs bosons $h^0$, $H^0$, $A^0$, and $H^\pm$
\cite{eberl:cit2,eberl:cit3}.
For the search of these particles a precise knowledge of all possible
decay modes is necessary.

The Higgs boson decays to supersymmetric (SUSY) particles could
be very
important if they are kinematically allowed. 
This can be the case for the charged Higgs boson $H^+$,
and the neutral Higgs bosons $H^0$ and $A^0$.
If all SUSY particles are very heavy, the $H^+$ decays dominantly
into $t\bar{b}$;
the decays $H^+ \to \tau^+ \nu$ and/or $H^+ \to W^+ h^0$ are
dominant below the $t\bar{b}$ threshold \cite{eberl:cit2,eberl:cit4}.
If all decay modes into SUSY particles are 
kinematically forbidden the $H^0$ and $A^0$ decay dominantly into a 
fermion pair of the third generation. In the case that the SUSY particles
are relatively light, the branching ratios of the $H^+$, $H^0$ and $A^0$
decays at
tree level were studied in \cite{eberl:citH+tree} and \cite{eberl:citH0tree}.
The SUSY--QCD corrections in ${\cal O}(\alpha_s)$ were calculated in the
on--shell scheme for the 
processes $H^+ \to t \bar b$ in \cite{eberl:citH+qcd,eberl:Jimenez},
$H^0, A^0 \to q \bar{q}$
in \cite{eberl:Coarasa}, and for the decays of 
all Higgs particles into squark pairs in \cite{eberl:citHiggsqcd}, including
squark--mixing and a proper renormalization of the mixing angle 
$\theta_{\tilde q}$
\cite{eberl:squarkprod}.\\
Within this work we will discuss the branching ratios of the Higgs decays 
including all SUSY--QCD corrections in ${\cal O}(\alpha_s)$. We will see
that the decay modes into
SUSY particles (squarks of the third generation, charginos
and neutralinos) become more important when the QCD corrections are taken
into account.
   
\section{The Tree Level}
We first review some tree--level results 
\cite{eberl:citH+tree,eberl:citH0tree,eberl:cit2}.
The squark mass matrix in
the basis ($\sq_L$, $\sq_R$),
with $\sq=\st$ or $\sb$, is given by \cite{eberl:cit2,eberl:cit3}
\begin{equation} \label{eberl:eq1}
\left( \begin{array}{cc}m_{LL}^2 & m_{LR}^2 \\ m_{RL}^2 & m_{RR}^2
\end{array} \right)=
(R^{\sq})^{\dagger}\left( \begin{array}{cc}m_{\sq_1}^2 & 0 \\ 0 &
m_{\sq_2}^2 \end{array}
\right)R^{\sq}\, ,
\end{equation}  
where
\begin{eqnarray} 
m_{LL}^2 &=& M_{\tilde{Q}}^2+m_q^2+m_Z^2\cos 2\beta
(I^{3L}_q-e_q\sin^2\theta_W), \\
m_{RR}^2 &=& M_{\{\tilde{U},
\tilde{D}\}}^2+m_q^2+m_Z^2\cos 2\beta e_q\sin^2
\theta_W, \\
m_{LR}^2=m_{RL}^2 &=& \left\{ \begin{array}{ll}
m_t(A_t-\mu\cot\beta) & (\sq=\st) \\
m_b(A_b-\mu\tan\beta) & (\sq=\sb) \end{array} \right. \, ,\label{eberl:eq4} 
\end{eqnarray}
and
\begin{equation}   \label{eberl:eq5}
R^{\sq}_{i\alpha}=\left(
\begin{array}{cc}\cos\theta_\sq & \sin\theta_\sq \\
-\sin\theta_\sq & \cos\theta_\sq \end{array}\right) \, .
\end{equation}   
The mass eigenstates $\sq_i(i=1,2)$ (with $m_{\sq_1}<m_{\sq_2}$) are
related to the
SU(2)$_L$ eigenstates $\sq_{\alpha}(\alpha=L,R)$ by
$\sq_i=R^{\sq}_{i\alpha}\sq_{\alpha}$.

\noindent The tree--level decay width
of $H^k \rightarrow \sq_i \bar{\sq}_j$ is then
given by 
\begin{equation} \label{eberl:eq6}
\Gamma^{tree}(H^k\rightarrow\sq_i\bar{\sq}_j) =
\frac{3 \kappa(m_{H^k}^2, m_{\sq_i}^2, m_{\sq_j}^2 )}{16\pi
m_{H^k}^3}|G_{ijk}^\sq|^2\,.
\end{equation}  
For $k = 1, 2, 3$ $H^k$ denotes the neutral Higgs bosons (i. e.
$H^1 \equiv h^0$,
$H^2 \equiv H^0$,
$H^3 \equiv A^0$)
and $\sq = \st, \sb$.
For $k = 4$ one has $H^4 \equiv H^+$ and  $\sq_i \equiv \st_i$,
$\sq_j \equiv \sb_j$, $(i,j = 1,2)$, and
$\kappa(x,y,z)\equiv
((x-y-z)^2-4yz)^{1/2}$.
The expressions for the couplings $G_{ijk}^\sq$ 
are given in \cite{eberl:citHiggsqcd}.
The decay widths of $H^+$ and $H^0$ into squarks can be large in the case of
large squark mixing. The decay width of $A^0$ into $\sq_1 \bar{\sq}_2$
is directly proportional to $|m_q (A_q c_q + \mu)|^2$ with $c_t = \cot\beta$ 
and $c_b = \tan\beta$. Starting  from the threshold  
these widths are steeply 
increasing with increasing $m_{H^k}$ 
up to a maximum and then decreasing.
For
large $m_{H^k}$ they become proportional to $1/m_{H^k}$.\\   
The decay widths into quarks are given by
 \begin{eqnarray}
\Gamma^{tree}(H^k\rightarrow q \bar q)  & = & \frac{3 g^2 m_q^2 
(d_k^q)^2 m_{H^k}}{32 \pi m_W^2 \sin^2\beta}\left(1 - \frac{4 
m_q^2}{m_{H^k}^2}\right)^{(3/2 - \delta_{k3})} \, , \quad (k = 1,2,3)\, , 
                         \\
\Gamma^{tree}(H^+\rightarrow t \bar b)  & = & 
\frac{3 \kappa(m_{H^+}^2, m_t^2, m_b^2)}{16 \pi m_{H^+}^3} \times
    \nonumber\\
&& \left[
(m_{H^+}^2 - m_t^2 - m_b^2) (y_t^2 + y_b^2) - 4 m_t m_b y_t y_b\right]\, ,
  \label{eberl:H+quarks}  
 \end{eqnarray}
with $d_1^t = - d_2^b = - \sin\alpha$, $d_2^t =  d_1^b = \cos\alpha$,
$d_3^t = \cos\beta$, $d_3^b = -\sin\beta$,
$\alpha$ being the $h^0$ -- $H^0$ mixing angle, and the Yukawa couplings $y_{t}$
and $y_{b}$:
\begin{equation}
   y_t = \frac{g}{\sqrt{2} m_W} m_t \cot\beta \, , \qquad 
   y_b = \frac{g}{\sqrt{2} m_W} m_b \tan\beta \, .	
    \label{eberl:yukawa}
\end{equation}
For large $\tan\beta$ $y_b$ can also become large and, therefore,
the sbottom and bottom modes become important.  
The formulae for the decay widths 
into charginos and neutralinos have essentially the same structure as 
eq.~(\ref{eberl:H+quarks}) with the 
appropriate masses and couplings (without the color 
factor 3), see e.~g. \cite{eberl:cit2}. If the mass of the decaying Higgs 
particle is large, the decay widths into fermions (quarks, 
charginos/neutralinos, \ldots) become proportional to $m_{H^k}$.\\
In the chargino/neutralino sector one has quite generally the following 
behaviour:\\
$M \ll |\mu|\,$: $\qquad$ ${\tilde\chi}_1$ is gaugino--like $\quad \to \quad$
$\Gamma^{tree}$ is small,\\
$M \gg |\mu|\,$: $\qquad$ ${\tilde\chi}_1$ is higgsino--like $\quad \to \quad$
$\Gamma^{tree}$ is large.\\

We have calculated the widths of all important modes of $H^+$,
$H^0$ and $A^0$ decays:\\
\begin{tabular}{lrcl} 
({\romannumeral 1}) &
$H^+$ &$\to$& 
$t \bar b$, $c \bar s$, $\tau^+ \nu_\tau$, $W^+ h^0$, $\st_i \bar\sb_j$, 
$\tilde{\chi}^+_k \tilde{\chi}^0_l$,  $\tilde{\tau}_i^+ 
\tilde{\nu}_\tau$, $\tilde{\ell}_L^+ \tilde{\nu}_\ell \,(\ell = e, \mu)$,\\
({\romannumeral 2})&
$H^0$ &$\to$& $t\bar{t}$, $b\bar{b}$, $c\bar{c}$,
$\tau^-\tau^+$, $W^+W^-$, $Z^0Z^0$, $h^0h^0$, $A^0A^0$,
$W^\pm H^\mp$, $Z^0A^0$,\\
&&& $\st_i\bar{\st}_j$,
$\sb_i\bar{\sb}_j$, $\sl^-_i\sl^+_j$, $\sn_\ell\bar{\sn}_\ell$
($\ell = e, \mu, \tau$), $\sw^+_i\sw^-_j$,
$\sz_k\sz_l$, and\\ 
({\romannumeral 3})&
$A^0$ &$\to$& $t\bar{t}$, $b\bar{b}$, $c\bar{c}$,
$\tau^-\tau^+$, $Z^0h^0$, $\st_1\bar{\st}_2$, $\st_2\bar{\st}_1$,
$\sb_1\bar{\sb}_2$, $\sb_2\bar{\sb}_1$,
$\stau^-_1\stau^+_2$, $\stau^-_2\stau^+_1$,\\
&&& $\sw^+_i\sw^-_j$, $\sz_k\sz_l$.
\end{tabular}

\vspace{2mm}

Formulae for these widths are found e.~g. in ref.\cite{eberl:cit2}.
In principle, also the decays $H^+ \to \tilde{u}_L \bar{\tilde d}_L, 
\tilde{c}_L \bar{\tilde s}_L$ and
$H^0 \to \sq_\alpha\bar{\sq}_\alpha$
($q = u, d, c, s$ and $\alpha = L, R$) could contribute
via their gauge couplings.
As the squarks of the first two generations are supposed
to be heavy, these decays will be strongly phase-space
suppressed.
Even if they were kinematically allowed, they would have a
rate at most comparable to that of $H^+ \to \tilde{\ell}^+ \sn_\ell$, 
$H^0$ $\to$ $\sl^-_i\sl^+_j$ and $\sn_\ell\bar{\sn}_\ell$
(see fig.~\ref{eberl:figbranch} below).
We have neglected loop induced decay modes (such as $H^+ \to W^+ Z^0, W^+ \gamma$,
$H^0 \to gg$,
and $\gamma\gamma$) and three-body decay modes \cite{eberl:gunion,eberl:zerwas1}.

\section{Numerical Results including SUSY--QCD corrections}
In the following branching ratios including SUSY--QCD corrections in
${\cal O}(\alpha_s)$ will be shown. For 
further details concerning the theoretical calculation of these corrections
we refer to \cite{eberl:citH+qcd,eberl:Coarasa,eberl:citHiggsqcd}.\\
We have chosen
\{$m_{A^0}$, $m_{t}$, $m_{b}$, $M$, $\mu$, $\tan\beta$, $M_{\tilde{Q}}$, $A$\}
as the
basic input parameters of the MSSM, taking
$M=(\alpha_2/\alpha_s(m_\sg))m_{\sg}=(3/5\tan^2\theta_W)M'$,
$M_{\tilde{Q}}(\st) : M_{\tilde{U}} : M_{\tilde{D}} : M_{\tilde{L}} :
M_{\tilde{E}} = 1 : \frac 8 9 : \frac{10}{9} : 1 : 1$ and
$A \equiv A_t = A_b = A_\tau$.
Here $M$ ($M'$) is
the SU(2) (U(1)) gaugino mass, $\alpha_2=g^2/4\pi$, and
($M_{\tilde{L}, \tilde{E}}$,
$A_{\tau}$)
are the mass matrix parameters of the slepton sector 
\cite{eberl:citH+tree,eberl:citH0tree}.
We have taken $m_t = 175$~GeV, $m_b = 5$~GeV, $m_Z = 91.2$~GeV, $m_W = 80$~GeV,
$\sin^2\theta_W=0.23$, $\alpha_2 = 0.0337$,
and $\alpha_s = \alpha_s(m_{H^k})$ for the $H^k$ decay. We have used
$\alpha_s(Q)=
12\pi/\{(33-2n_f)\ln(Q^2/\Lambda_{n_f}^2)\}$,
with $\alpha_s(m_Z)=0.12$, and the number of quark flavors $n_f=5(6)$ for
$m_b<Q\le m_t$ (for $Q>m_t$).\\
We have considered two scenarios:
\begin{center}
$\begin{array}{|c||c|c|c|c|c|}
\hline
\mbox{parameter set} & \tan\beta & M\,\mbox{[GeV]} & \mu\,\mbox{[GeV]}& 
M_{\tilde Q}\,\mbox{[GeV]}& A\,\mbox{[GeV]}\\
\hline
\mbox{I} & 2.5 & 160 & 350 & 95 & 300\\
\hline
\mbox{II} & 2.5 & 300 & -100 & 100 & -200\\
\hline
\end{array}$
\end{center}

\begin{figure}[h]
\begin{center}
\mbox{\epsfig{file=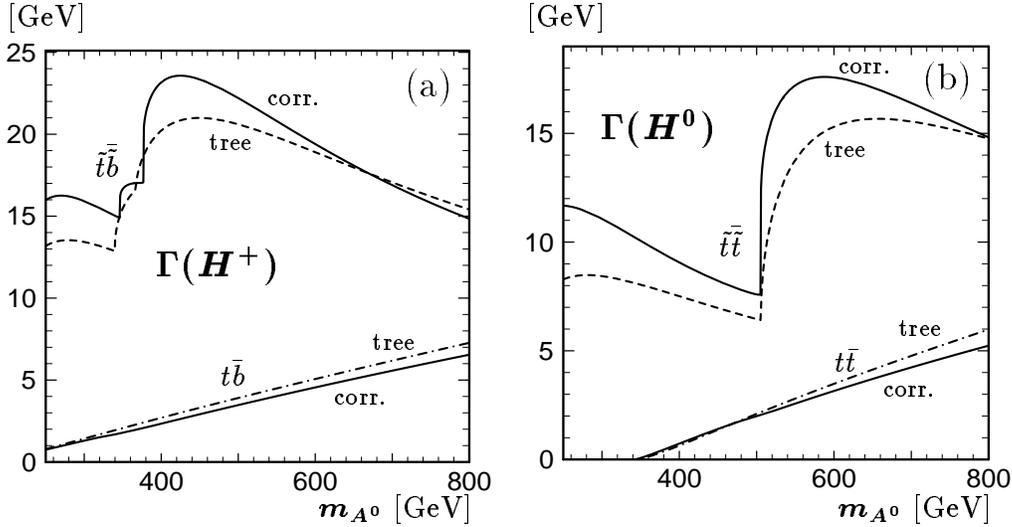,height=7cm,width=13.5cm}}
\end{center}
\caption[heberl1]{Tree--level and SUSY--QCD corrected decay widths} 
\label{eberl:figQCD}
\end{figure}

\begin{figure}
\begin{center}
\vspace*{-5mm}
\mbox{\epsfig{file=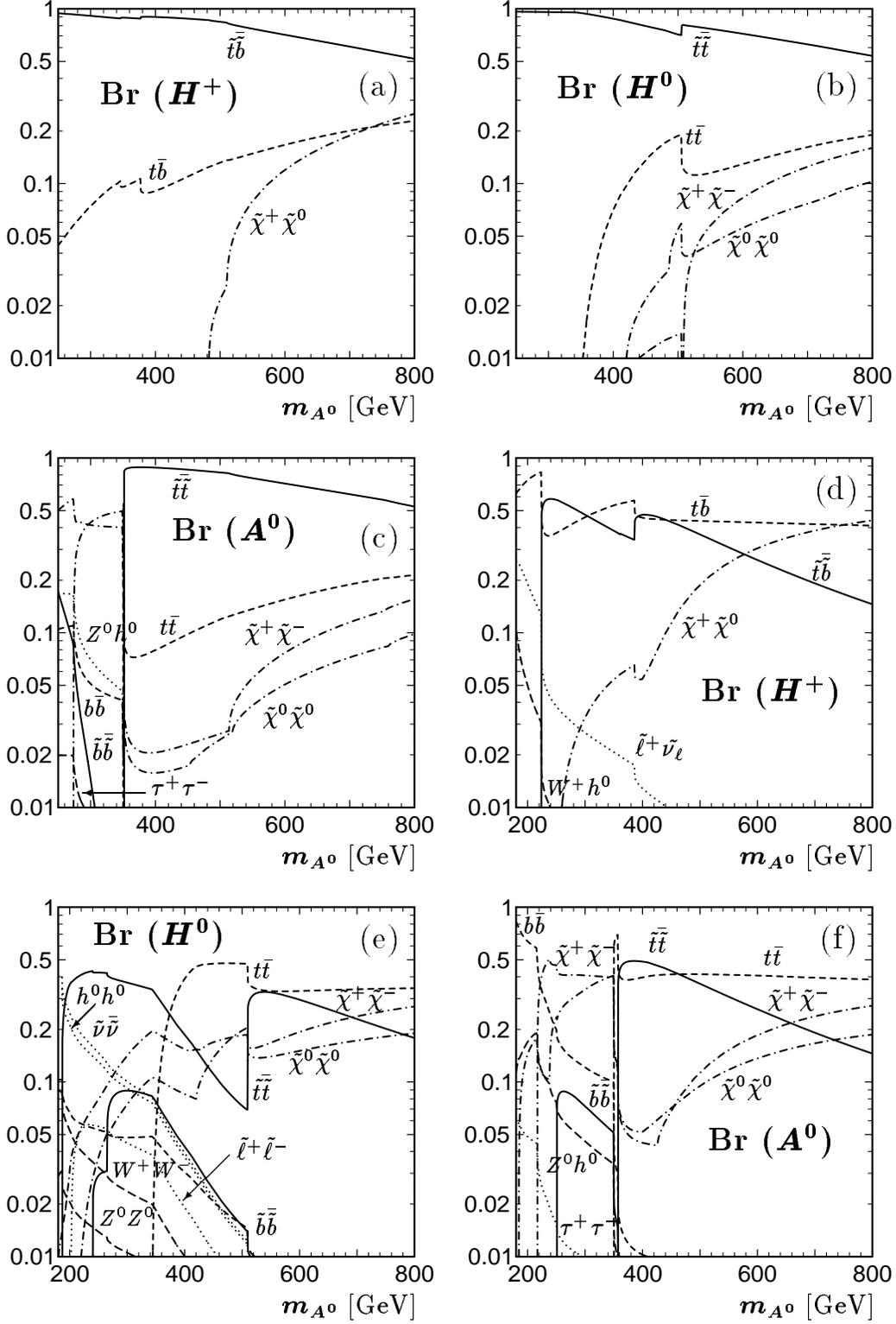,height=20.5cm,width=14cm}}
\end{center}
\vspace{-0.3cm}
\caption[heberl1]{Higgs particles branching ratios including SUSY--QCD corrections 
to quark and squark modes in 
${\cal O}(\alpha_s)$} 
\label{eberl:figbranch}
\end{figure}

We have implemented the new Higgs mass bound from ALEPH \cite{eberl:aleph},
$m_{h^0} \nge 70$~GeV for $\tan\beta = 2.5\,$. 
This leads for the parameter set~I to
$m_{A^0} \nge 250$~GeV and for the parameter set~II to $m_{A^0} \nge 180$~GeV.\\
For the parameter set~I we have (in GeV units)\\
$(m_{\st_1}, m_{\st_2}, m_{\sb_1}, m_{\sb_2},
 m_\sg, m_{\tilde{\chi}_1}^0,  m_{\tilde{\chi}_1}^+)$ =
(96, 255, 100, 130, 465, 73, 137),\\
and for the parameter set~II\\
$(m_{\st_1}, m_{\st_2}, m_{\sb_1}, m_{\sb_2},
 m_\sg, m_{\tilde{\chi}_1}^0,  m_{\tilde{\chi}_1}^+)$ =
(100, 257, 113, 137, 820, 92, 108).\\
Fig.~\ref{eberl:figQCD}~(a) shows the tree-level and the SUSY--QCD corrected
decay widths 
$\sum_{i,j} \Gamma(H^+ \to  \st_i \bar{\sb}_j)$ and
$\Gamma(H^+ \to  t \bar{b})$ and
Fig.~\ref{eberl:figQCD}~(b) shows the tree--level and the SUSY--QCD corrected 
decay widths 
$\sum_{i,j} \Gamma(H^0 \to  \st_i \bar{\st}_j)$ and
$\Gamma(H^0 \to  t \bar{t})$ using the parameter set~I.
The modes into bottom 
quarks and sbottoms are very small compared to the top and stop modes in
Fig.~\ref{eberl:figQCD}~(b) and therefore not shown.\\ 
Fig.~\ref{eberl:figbranch}~a--c show SUSY--QCD corrected branching ratios 
larger
than 1\% for the parameter set~I and Fig.~\ref{eberl:figbranch}~d--f for the 
parameter set~II.
All SUSY modes are summed up, e.~g. in the $H^0$ decay into
$\st \bar{\st} \equiv 
\sum_{i,j} \mbox{Br}(H^0 \to \st_i \bar{\st}_j)$.\\
In most cases, the SUSY--QCD corrections to the Higgs decays into quarks are
negative and into squarks positive. Therefore, the branching ratios for 
decays into squarks are enhanced by including the SUSY--QCD corrections.\\

\noindent{\bf Acknowledgements}\\
This work was supported by the "Fonds zur F\"orderung der 
wissenschaftlichen Forschung" of Austria, project no. P10843--PHY.

\nopagebreak

\end{document}